\documentclass{iaus}
\usepackage{graphicx}

\title[Chemical evolution of Local low mass systems] %% give here short title %% 
{About the Chemical Evolution of dSphs \\ 
(and the peculiar Globular Cluster $\omega$ Cen) }

\author[Andrea Marcolini \& Annibale D'Ercole]   %% give here short author list %%
{Andrea Marcolini$^1$
 \and Annibale D'Ercole$^2$}

\affiliation{$^1$ Centre for Astrophysics, University of Central Lancashire,
\\  Preston, Lancashire, PR1 2HE, United Kingdom
\\ email: {\tt amarcolini@uclan.ac.uk} \\[\affilskip]
$^2$ INAF, Osservatorio Astronomico di Bologna, \\
 via Ranzani 1, 40127 Bologna, Italy \\
email: {\tt annibale.dercole@bo.astro.it}} 

\pubyear{2008}
\volume{255}  %% insert here IAU Symposium No.
\pagerange{}
 \jname{Low-Metallicity Star Formation:\\
 From the First Stars to Dwarf Galaxies} 
\editors{L. Hunt, S.Madden \& R.Schneider, eds.}
\begin{document}

\maketitle

\begin{abstract}

 We present three dimensional hydrodynamical simulations aimed at
 studying the dynamical and chemical evolution of the interstellar
 medium (ISM) in isolated dwarf spheroidal galaxies (dSphs). This
 evolution is driven by the explosion of Type II and Type Ia
 supernovae, whose different contribution on both the dynamics and
 chemical enrichment is taken into account.  Radiative losses are
 effective in radiating away the huge amount of energy released by SNe
 explosions, and the dSph is able to retain most of the gas allowing a
 long period ($\ge 2-3$ Gyr) of star formation, as usually observed in
 this kind of galaxies.  We are able to reproduce the stellar
 metallicity distribution function (MDF) as well as the peculiar
 chemical properties of strongly O-depleted stars observed in several
 dSphs. The model also naturally predicts two different stellar
 populations, with an anti-correlation between [Fe/H] and velocity
 dispersion, similarly to what observed in the Sculptor and Fornax
 dSphs. These results derive from the inhomogeneous pollution of the
 SNe Ia, a distinctive characteristic of our model. We also applied
 the model to the peculiar globular cluster (GC) $\omega$ Cen in the
 hypothesis that it is the remnant of a formerly larger stellar
 system, possibly a dSph.

\end{abstract}

\firstsection % if your document starts with a section,
\section{Introduction}

Due to their proximity, the galaxies of the Local Group (see Mateo
1995 and Geisler et al. 2007 for a review) offer an unique opportunity
to study in details their structural, dynamical and chemical
properties and to test different theories of galaxy formation.
 
Owing to their low metallicity and lack of neutral hydrogen, it was
initially believed that dSphs are relatively simple objects whose ISM
is completely removed by SN II explosions after a very short intense
star formation period (e.g. Dekel \& Silk 1986).  Doubts about this
picture come from the high resolution spectroscopy of several dSphs
showing a wide range in metallicity. For istance, Shetrone et
al. (2001) have observed stars in Draco and Ursa Minor with values of
[Fe/H] in the range $-3\leq$[Fe/H]$\leq -1.5$. The same authors also
found that their observed dSphs have [$\alpha$/Fe] abundances that are
$0.2$ dex lower than those of Galactic halo field stars at the same
metallicity. This suggests that the bulk of the stars in these systems
formed in gas self polluted by SNe II as well as SNe Ia and that the
star formation (SF) must continue over a relatively long timescale in
order to allow a sufficient production of iron by SNe Ia (Ikuta \&
Arimoto 2002, Lanfranchi \& Matteucci 2004, Marcolini et al. 2006; but see
also Recchi et al. 2007, Salvadori et al. 2008).

A complex star formation history (SFH) is further suggested by several
facts: $i$) isolated low mass dSphs such as Phoenix (Young et
al. 2007) and Leo T (de Jong et al. 2008) were able to form stars up
to 100 Myr ago; $ii$) the SFHs of dwarf galaxies are strongly
dependent on their local environment, the fraction of passively
evolving galaxies dropping from $\sim$70\% in dense environments, to
zero in the rarefied field (Haines et al. 2007 ); $iii$) dwarf
ellipticals and dSphs cluster around the dominant spirals galaxies,
while gas rich star forming dwarf Irregulars are found at larger
distances (van den Bergh 1994). These points highlight the role of the
environment (tidal interaction/ram pressure stripping), and disfavors
a scenario in which the evolution is due uniquely to internal
processes.

\begin{figure}[]
% \vspace*{-2.0 cm}
\begin{center}
 \includegraphics[width=2.6in]{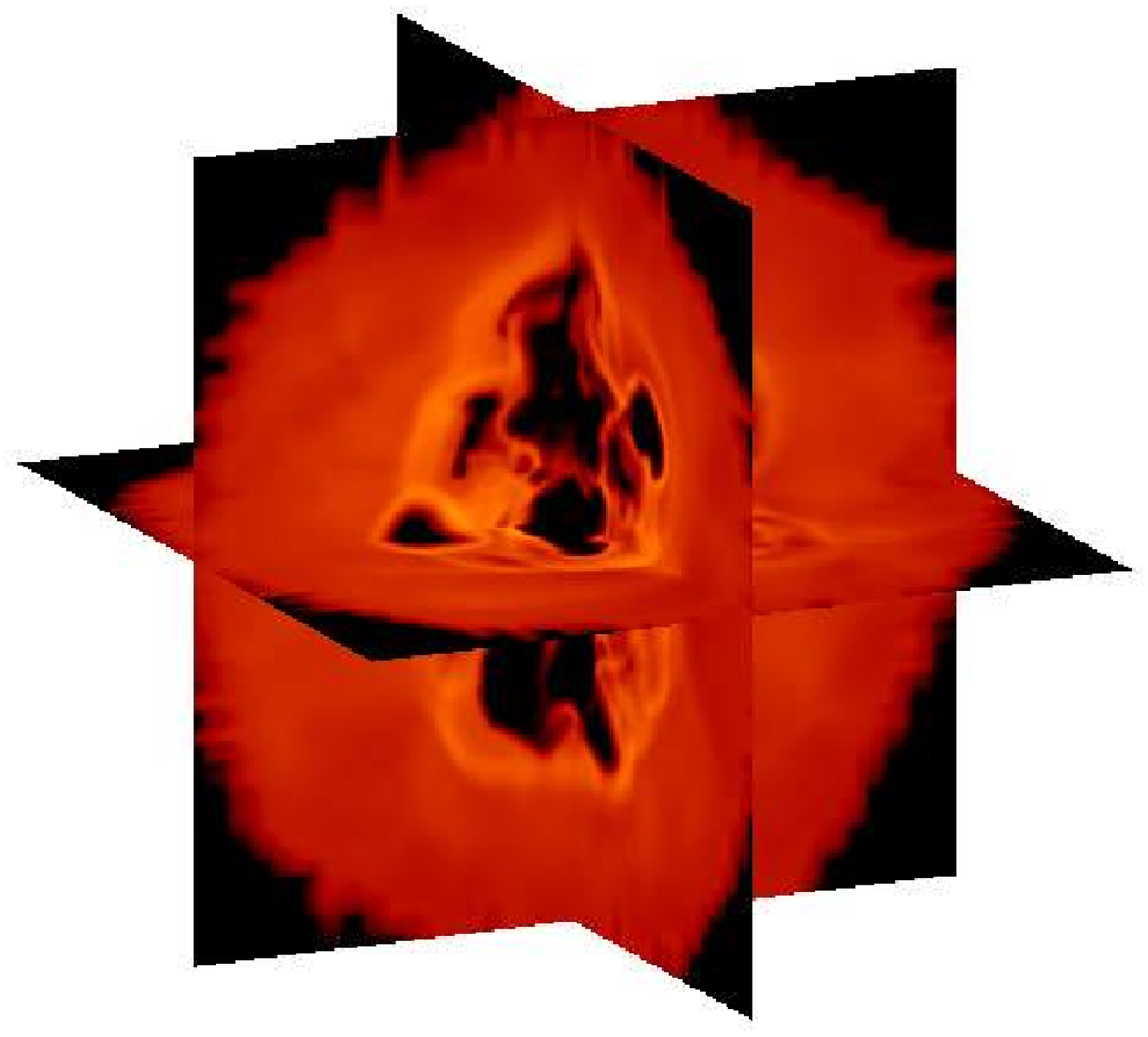}
  \includegraphics[width=2.6in]{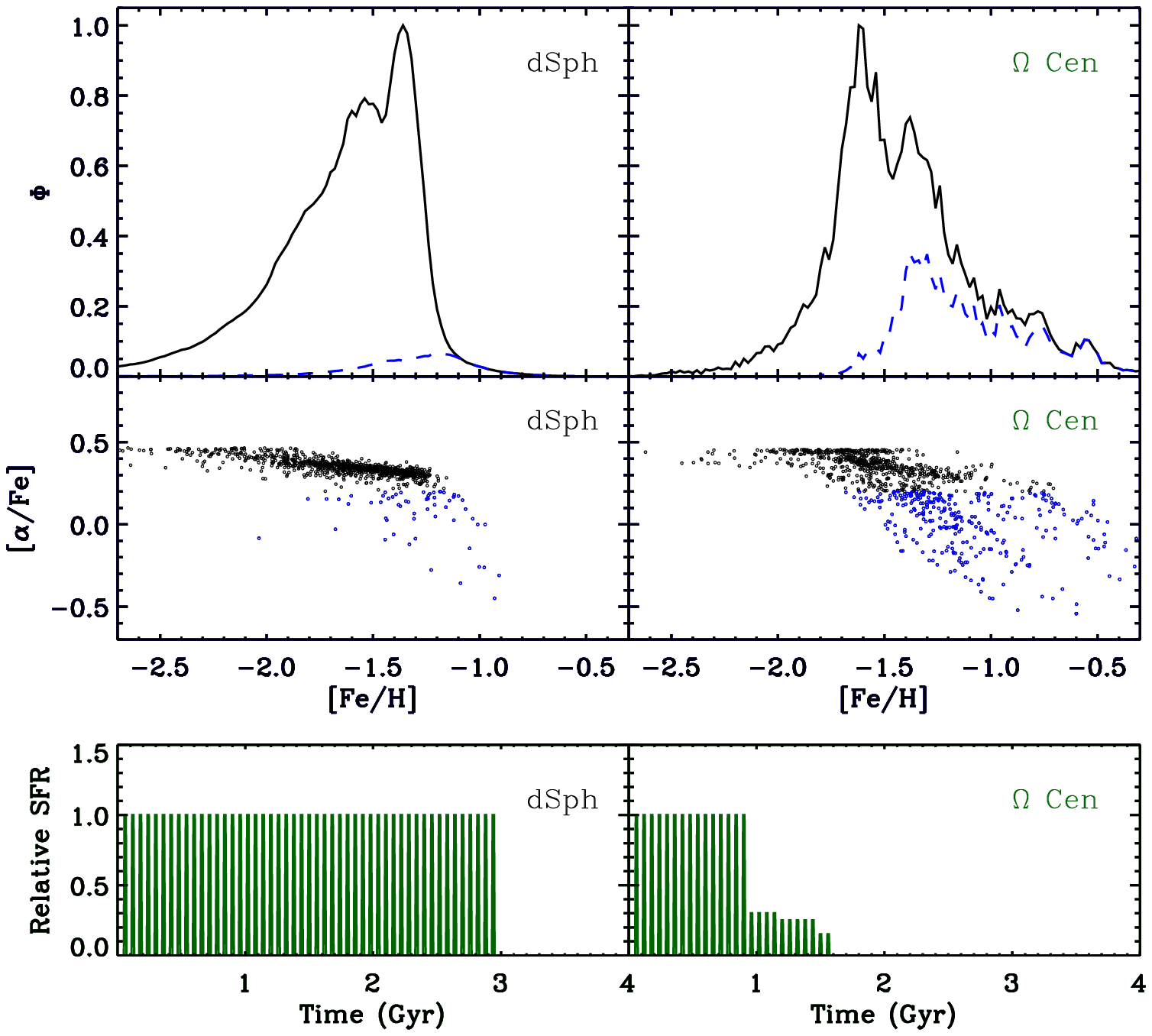}
% \vspace*{-1.0 cm}
  \caption{Left: distribution of the logarithm of the density in three
    orthogonal planes at $\sim$ 400 Myr after the beginning of the
    simulation. Note the inner bubble carved by the SNe explosions and
    the shocks propagating outward. Right: [Fe/H] distribution of the
    models resembling the Draco dSph (left panels) and the $\omega$ Cen
    globular cluster (right panels) together with the corresponding
    [$\alpha$/Fe]-[Fe/H] diagrams; the blue line and blue dots
    represent stars with [$\alpha$/Fe]$\le$0.2 (i.e. affected by SNe Ia
    inhomogenous pollution, see text). The lower panels represents the assumed
    SFH for the two models.}
   \label{fig1}
\end{center}
\end{figure}

Here we briefly report some results obtained by Marcolini et
al. (2006, 2007, 2008) with a model which turns to be consistent with
many properties of the Draco dwarf and, with minimal assumptions, to
the chemical properties of the peculiar system $\omega$ Centauri
(Marcolini et al. 2007), which is believed to be the remnant of an
ancient dSph.

\section{dSphs Model}

Let us consider the reference model of Marcolini et al. (2006) which,
although employed to study the general characteristics of dSphs, is
taylored to explore the evolution of the Draco dSph. The simulation
starts with the ISM in hydrostatic equilibrium in the extended dark
matter halo potential well. The amount of initial gas corresponds to
the cosmological baryonic fraction of the dark matter halo (M$_{\rm
ISM}=0.18$M$_{\rm DM}$). The SFH is given {\it a priori}, assuming
that stars form in a sequence of 50 instantaneous bursts separated in
time by 60 Myr (see Fig.~1). We also assume that SNe II explode at a
constant rate for 30 Myr after the occurrence of each starburst, while
the SNe Ia rate follows the prescription of Matteucci \& Recchi
(2001).  Each SN explosion is stochastically placed into the galaxy
according to its radial probability $P(r)=M_{\star}(r)/M_{\star,tot}$,
where $M_{\star}(r)$ and $M_{\star,tot}$ are the nowadays radial
stellar mass profile and total stellar mass, respectively
(cf. Marcolini et al. 2006 for more details).

\begin{figure}[]
% \vspace*{-2.0 cm}
\begin{center}
 \includegraphics[width=3.1in]{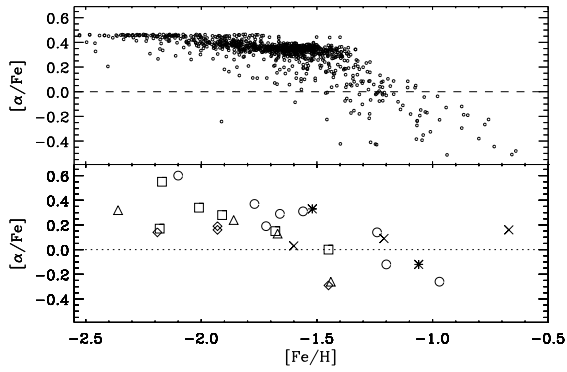}
% \vspace*{-1.0 cm}
 \caption{Abundance ratio [$\alpha$/Fe] versus [Fe/H] for 500 sampled
stars of the reference model at $t=2$ Gyr, compared to a dataset of 28
dSph stars (Draco: triangles ; Fornax: crosses; Leo I: asterisks;
Sextans: diamonds; Sculptor: circles; Ursa Minor: squares) collected
from the literature (see Marcolini et al 2008 for details).}
   \label{fig2}
\end{center}
\end{figure}

Although the total energy released by the SNe~II explosions is larger
than the binding energy of the ISM, efficient radiative losses enable
the galaxy to retain most of its gas, which thus remains available for
the aforementioned prolonged SFH.  The burst of SNe II associated with
each stellar burst pushes the bulk of the ISM to the outskirt of the
galaxy (see Fig.~1, left panel). Once the explosions cease
($\sim$30~Myr after each star burst episode), the ISM flows back
towards the center of the galaxy; when the next burst occurs, the gas
is pushed outwards again.  This oscillatory behavior leads to a rather
efficient and homogeneous pollution of the ISM by the SNe~II
ejecta. We note that as the SFH has been fixed {\it a priori}, it has
no direct relation to the gas reservoir within the galaxy. However,
stars do form during the quiescent periods between bursts, when the
gas has ``settled'', and an {\it a posteriori} consistency for the SFH
is recovered. For example a similar periodic ISM behaviour was
recovered in similar low-mass galaxies simulations performed by
Stinson et al. (2007). We point out that, while the observed dSphs are
deprived of gas, the galaxy in the present model can not expel the
bulk of its ISM by internal mechanisms. Following the discussion given
in the Introduction, an external cause, such as the interaction with
the Milky Way (Mayer et al. 2006), must be invoked at some evolutive
stage of the galaxy to get rid of its ISM.

Given their lower rate, SNe~Ia do not significantly affect the general
hydrodynamical behavior of the ISM, but their role is relevant for the
chemical evolution of the stars. Because of their longer evolutionary
timescales, SN~Ia progenitors created in previous starbursts continue
to explode during the subsequent quiescent periods, when the gas is
flowing back into the central region. During these periods the higher
ambient gas density (together with the lower SNe~Ia explosion rate)
cause the SNe~Ia remnants to be isolated from one another, forming
chemically inhomogeneous regions (we refer to these regions as ``SNe
Ia pockets''). These pockets are ``washed out'' by successive phases
of expansion and collapse of the ISM, due to the effects of SNe~II,
but new pockets form during the quiescent phases between consecutive
starbursts. At odd with the ejecta of SNe~II, SN~Ia debris are rich in
iron and deficient in $\alpha$-elements. Thus, stars forming in the
SN~Ia pockets possess lower [$\alpha$/Fe] and higher [Fe/H] ratios
than those formed elsewhere.

\subsection{Comparison with Local dSphs}

As stressed above, the reference model discussed in Marcolini et
al. (2006) was tailored to fit the Draco galaxy. In Fig.~1 we show the
MDF of this model at the end of the simulation (3 Gyr). The maximum
value of the distribution occurring at [Fe/H]$\sim$-1.5 and the mean
value of $\langle$[Fe/H]$\rangle$=-1.65 with a spread of $\sim$1.5 dex
are compatible with observations (Bellazzini et al. 2002). The
agreement is larger at the high metallicity tail of the distribution
which, in our model, is shaped by the stars formed in the SN Ia
pockets.

As discussed above, the SNe Ia are also responsible of the
[$\alpha$/Fe] spread present at larger values of [Fe/H] in the
observed dSphs, as shown in Fig. 2 (lower panel). From this figure it
is also apparent that the most metal poor stars have
[$\alpha$/Fe]$\simeq0.5$, typical of pure SN II enrichment, while a
decrement of $\sim 0.2$ dex in the plateau is achieved at higher
metallicities.  This drop occurs in our simulations after $t\sim
2.0-3.0$ Gyr to allow the cumulative effect of SNe Ia to be
appreciable (top panel of Fig.~2). We thus conclude that, at least in
our models, only a prolonged star formation history ($> 2.0$ Gyr) can
account for the chemical differences between the Galactic halo and the
dSphs as shown by Shetrone et al. (2001). Such a long SFH for dSphs is
also found in the chemical models by Fenner et al. (2006) who are
unable to reproduce the Ba/Y ratio unless stars formed over an
interval long enough for the low-mass stars to pollute the ISM with
$s$-elements.

The inhomogeneous pollution by SNe Ia discussed above is particularly
important in the central galactic region, where the SN~Ia rate is
greater, and the density of the ambient gas is higher. This naturally
accounts for a radial segregation of Fe-rich stars in the central
regions of dSphs. As these stars in our model are also
$\alpha$-depleted, a similar radial segregation of $\alpha$-poor stars
should be observed to test our model. We finally stress that our model
envisages a central depression in the radial distribution of the
stellar velocity dispersion (Marcolini et al. 2008). This naturally
entails two different stellar populations with an anti-correlation
between [Fe/H] and velocity dispersion, which has been observed in the
case of the Sculptor dSph (Tolstoy et al. 2004) and the Fornax dSph
(Battaglia et al. 2006).

\begin{figure}[]
% \vspace*{-2.0 cm}
\begin{center}
 \includegraphics[width=3.5in,height=1.9in]{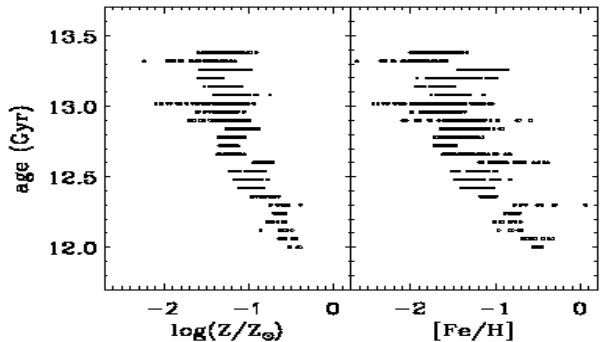}
% \vspace*{-1.0 cm}
 \caption{Age-Z and age-[Fe/H] distributions of 1000 sampled stars for
the $\omega$ Cen model. Note the metallicity spread among coeval
stars.}
   \label{fig3}
\end{center}
\end{figure}

\subsection{$\omega$ Cen}

The stellar system $\omega$~Cen (NGC~5139) is unique among Galactic
star clusters in terms of its structure, kinematics, and stellar
content. It is the only known GC showing a clear [Fe/H] spread
spanning the metallicity range $-1.6<$[Fe/H]$<-0.6$ (Norris et
al. 1996, Johnson et al. 2008). Recent photometric surveys have
revealed the presence of multiple sequences in its color-magnitude
diagram (CMD), indicating a complex star formation history
(e.g. Sollima et al. 2005). It is suggested that $\omega$~Cen is the
nucleus of a larger stellar system, possibly a dwarf galaxy, that lost
most of its stars and gas in the interaction with the Milky Way
$\sim$10~Gyr ago (e.g. Bekki \& Norris 2006).  In this framework
Marcolini et al. (2007) focused on the evolution of the central region
of their dSph model, where the inhomogeneous pollution of the SNe Ia
is particularly effective.

In Fig.~1 the MDF within the inner 90 pc is plotted assuming a total
SFH lasting $\sim$ 1.5 Gyr as shown in the same Figure (see Marcolini
et al. 2007 for more details and about the assumptions used for this
model). The MDF shows a bimodal structure similar to that observed in
$\omega$~Cen (e.g. Norris et al. 1996), with a maximum at
[Fe/H]=$-1.6$ and a secondary peak at [Fe/H]$\sim-1.3$ accounting for
$\sim25$\% of the cluster's stellar content.  Comparing the
[$\alpha$/Fe]-[Fe/H] diagram of this model with the one typical of
dSphs (c.f. Fig. 1), it is possible to note the much larger number of
$\alpha$-depleted stars (see also the blue line in the MDF which
represents stars with $\alpha$/Fe$\le$0.2).  The diagram is in
reasonable agreement with the findings of Pancino et al. (2002) and
Origlia et al. (2003), who find that while the metal-poor and
intermediate-metallicity stellar populations of $\omega$~Cen have the
expected $\alpha$-element overabundance observed in halo and GC stars
($\langle$[$\alpha$/Fe]$\rangle$ $\simeq$ 0.3-0.4), the most
metal-rich population ([Fe/H]$\sim -0.6$) shows a significantly lower
$\alpha$-enhancement ($\langle$[$\alpha$/Fe]$\rangle \simeq$0.1).

Given the distictive role of the SN Ia pollution, the [Fe/H] content
of this system is not simply proportional to the metal content and a
large spread in the age-metallicity relation is always present (see
Fig.~3). These peculiarities have important consequences on the
cluster CMD. In fact, they reduce by a factor of 40\% the large (and
still unexplained) helium overabundance usually invoked to account for
the anomalous position of the blue main sequence observed $\omega$
Cen.

%\begin{discussion}

%\discuss{Massey}{Im wondering if you have considered the expected intrinsic dispersion in absolute
%magnitude of WRs -- if you consider the (large) mass range that becomes an
%early WN or late WC according to the evolutionary models, wouldnt you expect a large
%dispersion in M$_v$?}

%\discuss{van der Hucht}{Indeed, we will be always left with some intrinsic scatter in M$_v$ due
%to mass differences within the same spectral subtype. But in my opinion, the current
%large dispersion is for a large fraction due to incertainties of the adopted distances of
%open clusters and OB associations.}

%\end{discussion}


\begin{thebibliography}{}

\bibitem[Battaglia et al.]{battaglia2006} {Battaglia G. et al., 2006,
A\&A, 459, 423}

\bibitem[Bekki 
\& Norris(2006)]{beknor06} Bekki, K., \& Norris, J.~E., 2006, ApJL, 637, L109 

\bibitem[{Bellazzini} {et~al.}(2002)]{bellazzini2002} {Bellazzini}
M., {Ferraro} F.~R., {Origlia} L., {Pancino} E., {Monaco} L., {Oliva}
E., 2002, AJ, 124, 3222

\bibitem[{Dekel} \& {Silk} {1986}]{dekel1986} {Dekel} A., {Silk} J.,
 1986, ApJ, 303, 39

\bibitem[{de Jong}, et al. {2008}]{dejong2008} {de Jong} et al., 2008,
astro-ph/08014027

\bibitem[{Fenner}, Y. and {Gibson}, B.~K. and {Gallino}, R. and
{Lugaro}, M.]{fenner2006}{Fenner}, Y., {Gibson}, B.~K., {Gallino}, R.,
{Lugaro}, M., 2006, ApJ, 646, 184	

\bibitem[{Geisler}, D., {Wallerstein}, G., {Smith}, V.~V., 
{Casetti-Dinescu}, D.~I.]{geisler2007} {Geisler}, D.,
{Wallerstein}, G., {Smith}, V.~V., {Casetti-Dinescu}, D.~I.,
2007, PASP, 119, 939

\bibitem[{Haines} et~al. {2007}]{haines2007} {Haines} C.,
{Gargiulo} A., {La Barbera} F., {Mercurio} A., {Merluzzi},
{Busarello}, 2007, MNRAS, 381, 7

\bibitem[{Ikuta}, C. and {Arimoto}, N.]{ikuta2002}{Ikuta}, C. and
{Arimoto}, N., 2002, A\&A, 391, 55

\bibitem[Johnson et al. (2008)]{johnson2008}{Johnson}, C.~I. and
{Pilachowski}, C.~A. and {Simmerer}, J. and {Schwenk}, D.,
astro-ph/08042607

\bibitem[{Lanfranchi} \& {Matteucci},{2004}]{lanfranchi2004}
{Lanfranchi} G.~A.,  {Matteucci} F.,  2004, MNRAS, 351, 1338

%\bibitem[{Marcolini} {et~al.}(2003)]{marcolini2003} {Marcolini} A.,
%{Brighenti} F., {D'Ercole} A., 2003, MNRAS, 345, 1329

\bibitem[Marcolini et al. (2006)]{marcolini2006} {Marcolini A.,
D'Ercole A., Brighenti F., Recchi S., 2006, MNRAS, 371, 643}

\bibitem[Marcolini et al. (2007)]{marcolini2007} {Marcolini A.,
Sollima A., D'Ercole A., Gibson B.K., Ferraro F.R., 2007, MNRAS, 382, 443}

\bibitem[Marcolini et al. (2008)]{marcolini2008} {Marcolini A.,
D'Ercole A., Battaglia G., Gibson B.K, 2008, MNRAS, 386, 2173}

\bibitem[Mateo 1998]{mateo1998}{Mateo M.~L.,  1998, ARA\&A, 36, 435}

\bibitem[{Matteucci}, F. and {Recchi}, S.]{matteucci2001}{Matteucci},
F. and {Recchi}, S., 2001, ApJ, 558, 351

\bibitem[{Mayer} {et~al.}(2006)]{mayer2006} {Mayer} L.,
{Mastropietro} C., {Wadsley} J., {Stadel} J., {Moore} B., 2006, MNRAS,
369, 1021

\bibitem[{Norris} {et~al.}]{norris1996} {Norris} J.~E., {Freeman}
K.~C., {Mighell} K.~J., 1996, ApJ, 462, 241

\bibitem[{Origlia} {et~al.}(2003)]{origlia2003} {Origlia} L.,
{Ferraro} F.~R., {Bellazzini} M., {Pancino} E., 2003, ApJ, 591, 916

%\bibitem[{Pancino} {et~al.}(2000)]{pancino2000} {Pancino} E., {Ferraro}
%F.~R., {Bellazzini} M., {Piotto} G., {Zoccali} M., 2000, ApJ, 534, L83

\bibitem[{Pancino} {et~al.}(2002)]{pancino2002} {Pancino} E.,
{Pasquini} L., {Hill} V., {Ferraro} F.~R., {Bellazzini} M., 2002, ApJ,
568, L101

\bibitem[{Recchi}, S. and {Theis}, C. and {Kroupa}, P. and {Hensler},
G.]{recchi2007}{Recchi}, S. and {Theis}, C. and {Kroupa}, P. and {Hensler},
G, 2007, A\%A, 470, 5

%\bibitem[{Ricotti} (2008)]{ricotti2008}{Ricotti} M., 2008,
%astro-ph/08062402


\bibitem[{Salvadori}, S. and {Ferrara}, A. and {Schneider}, R.
(2008)]{salvadori2008}{Salvadori}, S. and {Ferrara}, A. and
{Schneider}, R., 2008, MNRAS, 386, 348


\bibitem[{Shetrone} et~al. {2001}]{shetrone2001} {Shetrone} M.~D.,
{C{\^o}t{\'e}} P., {Sargent} W.~L.~W., 2001, ApJ, 548, 592

%\bibitem[{Shetrone} et~al. {2003}]{shetrone2003} {Shetrone} M., {Venn}
%K.~A., {Tolstoy} E., {Primas} F., {Hill} V., {Kaufer} A., 2003, AJ,
%125, 684

\bibitem[Sollima et~al. (2005)]{sollima2005} {Sollima} A., {Ferraro}
F.~R., {Pancino} E., {Bellazzini} M., 2005, MNRAS, 357, 265

\bibitem[{Stinson}, G.~S., {Dalcanton}, J.~J., {Quinn}, T.,
{Kaufmann}, T., {Wadsley}, J.]{stintson2007}{Stinson}, G.~S.,
{Dalcanton}, J.~J., {Quinn}, T., {Kaufmann}, T., {Wadsley},
J., 2007, ApJ, 667, 170

\bibitem[Tolstoy \etal\ (2004)]{tolstoy2004}{Tolstoy E., et al. 2004,
ApJ, 617, L119}

\bibitem[]{} van den Bergh S., 1994, ApJ, 428, 617

\bibitem[{Young} et~al. (2007)]{young2007} {Young} L.~M., {Skillman}
E.~D., {Weisz} D.~R., {Dolphin} A.~E., 2007, ApJ, 659, 331

\end{thebibliography}
\end{document}